\documentclass[twocolumn,twoside,slac_two]{revtex4}
\usepackage{graphicx}

\usepackage{amsmath,amssymb}

\usepackage{fancyhdr}
\usepackage{longtable}

\begin{document}
\title{Hadronic- and electromagnetic-cores of air-showers observed by hybrid experiments at high mountains}
       
\author{M.Tamada}
\affiliation{Faculty of Science and Engineering, Kinki University, 
Higashi-Osaka, 577-8502 Japan}

\begin{abstract}
Characteristics of the high energy families (bundle of high energy e,$\gamma$) and
hadrons in the air-showers  detected in the hybrid
experiment together with emulsion chamber and AS-array at Mt.Chacaltaya are studied in detail 
by comparing with those of CORSIKA simulations
using interaction models of QGSJET and EPOS.
Because the atmospheric families and hadron component have more direct information of the nuclear interaction, 
correlations between atmospheric families and  burst (hadron component of air-showers) accompanied
to air-showers are more sensitive to the mechanism of the the cosmic-ray interactions.
The burst size dependence of the family energy is compared with those of simulations.
It is found that the family energy accompanied by the air-showers with the larger burst-size is 
systematically smaller than that expected in the simulated events.
The experimental results can not be described simply by changing the chemical
composition of primary cosmic-rays and this indicates  that the x-distribution of secondary
particles in  cosmic-ray interactions becomes much steeper than that assumed
in the simulation models.

\vspace{1pc}
\end{abstract}

\maketitle

\thispagestyle{fancy}


\section{Introduction}

The main interests of the cosmic-ray study  now shift to the highest energy
region, $E_0 \ge 10^{19}$ eV, using huge experimental
apparatus to search for astrophysical sources  and the acceleration or emission mechanism of 
those extreme high energy cosmic-rays.
However,  half a century after its discovery, the ``knee'' in the cosmic-ray
spectrum, a steepening of the energy spectrum at $10^{15} \sim 10^{16}$ eV,
 is still not well understood.
Because of the low intensities, direct observations of primary cosmic-rays in this energy range
 are still not possible and  so  various types of air-shower experiments at high mountains and 
 also at ground level have been carried out in order to investigate the  chemical composition of primary
 cosmic-rays in this energy region which gives important information on the physical origin of cosmic-rays \cite{SYS96,SYS00,Tibet00,TienShan,BASJE,Kascade,AP18_Swordy}.
The experimental data in those indirect measurement are usually interpreted by 
comparing with Monte Carlo simulations assuming some models of cosmic-ray interactions.
Many experimental groups claim that the fraction of heavy
primaries increases rapidly beyond the ``knee'' region, e.g.,
 the fraction of protons is estimated 
by the Tibet AS-$\gamma$ group  \cite{Tibet06} to be 
as small as $\sim$10\% of
all particles for $E_0= 10^{15} - 10^{16}$ eV.
The results, however, depend on their assumed interaction model.
For example, the  EPOS model recently proposed \cite{EPOS06,EPOS07} 
gives muon numbers much more than the QGSJET model. 
The events which can be interpreted due to heavy primary when we employ the 
QGSJET model as nuclear interactions are interpreted due to proton primaries
when EPOS is used as the interaction model \cite{H.Ulrich}.
Thus the interpretations  rely heavily on the Monte Carlo  calculations.
In fact, various experimental groups give various data on chemical
composition in this energy region and the results are still very confusing.
We should examine whether the overall experimental data can be
well interpreted by the assumed model before drawing a conclusion.

Hybrid experiments operating simultaneously an air-shower array,
a hadron calorimeter and an emulsion chamber
have been carried out  at Mt. Chacaltaya  (5200m, Bolivia) \cite{SYS96,SYS00},
 Yang-bajing (4300m, China) \cite{Tibet00,Tibet06}
and Tien-Shan (3340m, Kazakhstan) \cite{TienShan} for studying 
cosmic-ray nuclear interaction in the energy region around
$10^{15} - 10^{17}$ eV.
In the hybrid experiments, we can obtain the air-shower size, $N_e$, from the 
air-shower array data, particle-density, $n_b$, which are closely connected to
the hadron component in the air-shower, from the hadron calorimeter (burst detector)
and energy and geometrical position of individual high energy electromagnetic particles by the emulsion chamber.
Correlations between air-showers  and accompanying families
were studied so far in these three experiments by comparing experimental data and simulated
data \cite{SYS_Merida,icrc09_tama}.
In the present paper we show some results obtained by studying  correlations  between the 
hadron component (data of hadron calorimeters) and families observed by emulsion
chamber which are considered to be very sensitive to the mechanism of cosmic-ray
interactions, using data of the Chacaltaya hybrid experiment.

\begin{figure}
\includegraphics[width=6cm]{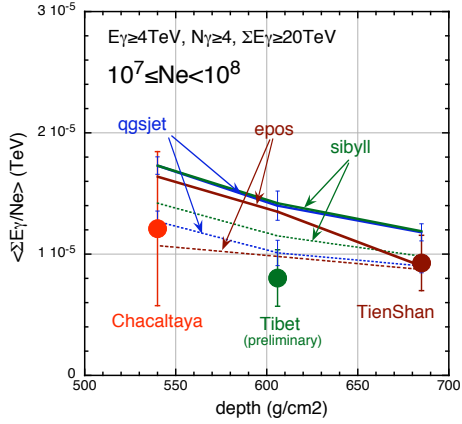}
\caption{\small 
Depth dependence on the average family energy normalized by air-shower size for 
events with $10^7 \le N_e < 10^8$. Solid lines are for proton-dominant composition
and dotted lines for heavy-dominant composition\cite{icrc09_tama}.}
\label{fig1}
\end{figure}
\begin{figure}
\includegraphics[width=6cm]{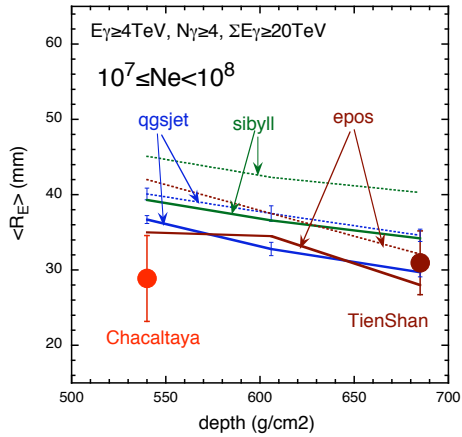}
\caption{\small 
Depth dependence on the average lateral spread of EAS-triggered families 
with $10^7 \le N_e < 10^8 $\cite{icrc09_tama}. }
\label{fig2}
\end{figure}

\section{Short summary of the analysis on  air-showers and
accompanied families}

The shower-size, $N_e$, dependence on the family
energy and on the lateral spread of the showers in the air-shower-triggered families were
studied in the three hybrid experiments at high mountains.
In Ref.\cite{icrc09_tama} we have shown that the average family energy normalized 
by associating air-shower size,
$\Sigma E_{\gamma} / N_e$, of the events with  $N_e \ge 10^7$ observed in these hybrid experiments
agree more or less to those expected in the case of heavy-dominant composition of
primary particles, as  shown in Figure~\ref{fig1}, though the difference in the average value between 
the two chemical composition,
proton-dominant and heavy-dominant, is not clear at the Tien-Shan  altitude, especially in the case of the EPOS model.
Some details of simulations are shown in Ref.\cite{icrc09_tama}.
The lateral spread of high energy showers in the families accompanied by the air-showers with  $N_e \ge 10^7$
was also studied\footnote{
There is no official publication about the lateral spread of families accompanied by air-showers
in Tibet $AS\gamma$ experiments.}.
The average lateral spread of showers in those families  observed in the  Chacaltaya experiment, shown in Figure~\ref{fig2}, 
is found to be smaller than that expected in the case of a heavy-dominant composition.
The difference in the average lateral spread between the two chemical compositions is again not clear
in the Tien-Shan data.
The Chacaltaya data show that the proton-dominant composition is favorable to explain the small lateral spread of 
the families but the heavy-dominant composition is favorable to explain small family energy.
Thus the increase of heavier composition of primary cosmic rays 
alone can not explain the general characteristics of 
air-shower-triggered families, contrary to the results of the Tibet group and others \cite{BASJE,Tibet06,Horandel03}. 
\\
\indent
The data of hadron calorimeters were also analyzed by the Chacaltaya and Tibet groups.
The Tibet group concluded from the analysis that the experimental data
were well explained by a heavy dominant composition of primary particles \cite{Tibet00} but the 
Chacaltaya group concluded that the number of hadrons in the air-showers was
less than expected \cite{SYS00}.
\\
\begin{figure}
\begin{center}
\includegraphics[width=6cm]{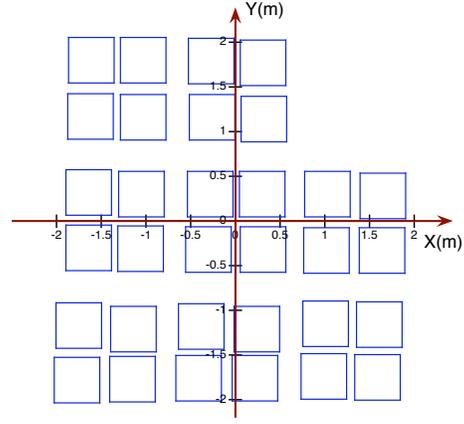}
\end{center}
\caption{\small 
Configuration of 32 blocks of emulsion chambers and hadron calorimeters
at the center of the Chacaltaya air-shower array.}
\label{fig3}
\end{figure}
\begin{figure}
\begin{center}
\includegraphics[width=6cm]{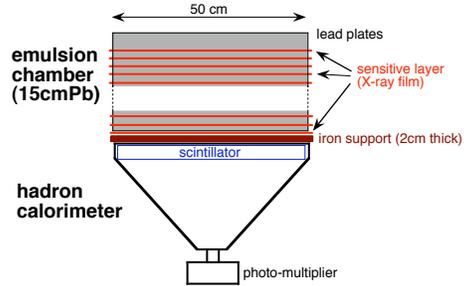}
\caption{The 
structure (side view) of one  of the  blocks of the emulsion chamber 
and the hadron calorimeter (burst detector).
}
\label{fig4} 
\end{center}
\end{figure}

\section{Hybrid experiment at Mt.Chacaltaya}

 The air-shower array covers a circular area within a radius about 
50 m  by 35 plastic scintillation detectors to measure
the lateral distribution of the electron density of the air-showers.
In the center of the air-shower array, 32 blocks of emulsion chambers (0.25 m$^2$ each) are installed
 (see Figure~\ref{fig3}). Each block of the emulsion
chamber consists of 30 lead plates each of 0.5 cm thick and
14 sensitive layers of X-ray film which are inserted after every
1 cm lead. The total area of the emulsion chambers is 8 m$^2$.
Hadron calorimeters with plastic scintillator of 5cm thickness are installed 
underneath the respective blocks of the emulsion chamber (see Figure~\ref{fig4}).
An iron support of 2 cm thick is inserted between the emulsion chamber
and the hadron calorimeter.
Some detail of the Chacaltaya hybrid experiment are described in 
Refs.\cite{SYS96, SYS00}

\section{Simulations}
\subsection{Air-showers and families}
For generating extensive air-showers and families
we use the CORSIKA simulation code (version 6.735) \cite{CORSIKA} 
employing the QGSJET  model (QGSJET01c) \cite{QGSJET} and the 
 EPOS model  (EPOS 1.60) \cite{EPOS06,EPOS07} 
 for the cosmic-ray nuclear interaction.
Primary particles of $E_0 \ge 10^{15}$ eV are sampled respectively
 from the power law energy spectrum of integral power index $-1.7$, for pure protons
 and pure iron,  and also from the energy 
spectrum of primary cosmic rays with proton dominant and a heavy dominant 
chemical composition. Some details of the chemical composition are  shown in Table 1.
The thinning energy is fixed to be 1 GeV.
Shower size, $N_e$, at the observation point is calculated  
using the NKG option in the simulation. 
The air-shower center is randomly sampled within an area of $\pm 2.5$ m
in the X and Y directions from the center of the hadron detectors (see Figure~\ref{fig3}).

\begin{table*}[htb]
\begin{center}
\footnotesize
\caption{\small Chemical composition of primary cosmic-rays and air-showers}
\begin{tabular}{@{\hspace{\tabcolsep}%
\extracolsep{\fill}}lccccc|ccccc} \hline
 & \multicolumn{10}{c}{sampled primary particles} \\
 & \multicolumn{5}{c}{proton dominant} & \multicolumn{5}{c}{heavy dominant}\\
$E_0$ (eV) & protons & He & CNO & heavy & Fe  
& protons & He & CNO & heavy & Fe \\ \hline
$10^{15} - 10^{16}$ & 42 \% & 16 \% & 16 \% & 14 \% & 12 \% 
                    & 17 \% & 10 \% & 18 \% & 15 \% & 40 \% \\
$10^{16} - 10^{17}$ & 42 \% & 12 \% & 13 \% & 15 \% & 18 \% 
                    & 14 \% & 8 \% & 17 \% & 14 \% & 47 \% \\
\hline \hline
\multicolumn{11}{c}{air-showers accompanied by burst (CORSIKA/QGSJET)}\\
        & protons & He & CNO & heavy & Fe 
& protons & He & CNO & heavy & Fe \\ \hline
 $10^6 < Ne < 10^7$ & 57 \% & 18 \% & 11 \% & 9 \% & 5 \% 
  & 31 \% & 13 \% & 11 \% & 13 \% & 26 \% \\
$10^7 < Ne < 10^8$  & 46 \% & 11 \% & 13 \% & 11 \% & 18 \% 
           & 16 \% & 12 \% & 21 \% & 9 \% & 42 \% \\ \hline
\multicolumn{11}{c}{air-showers accompanied by families of $\Sigma E_{\gamma} \ge 10$ TeV
 (CORSIKA/QGSJET)}\\
        & protons & He & CNO & heavy & Fe 
& protons & He & CNO & heavy & Fe \\ \hline
 $10^6 < Ne < 10^7$ & 70 \% & 19 \% & 6 \% & 4 \% & 2 \% 
  & 48 \% & 17 \% & 17 \% & 8 \% & 10 \% \\
$10^7 < Ne < 10^8$  & 50 \% & 9 \% & 14 \% & 7 \% & 20 \% 
           & 23 \% & 10 \% & 19 \% & 3 \% & 45 \% \\
\hline
\end{tabular}
\end{center}
\end{table*}

\begin{figure}
\begin{center}
\includegraphics[width=5cm]{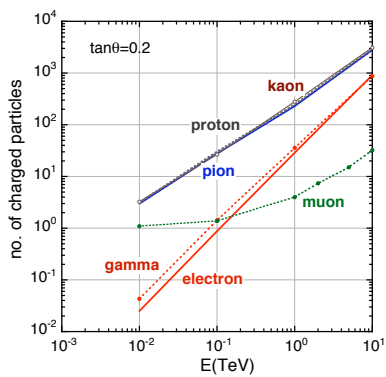}
\caption{\small 
Energy dependence of average number of charged particles arriving to the
scintillator of the hadron calorimeter for $\pi^- $, proton, $K^-$, $\mu^-$, e$^-$ and
$\gamma$ incidence.
}
\label{fig5}
\end{center}
\end{figure}
\begin{figure}
\begin{minipage}{0.2\textwidth}
\includegraphics[width=4cm]{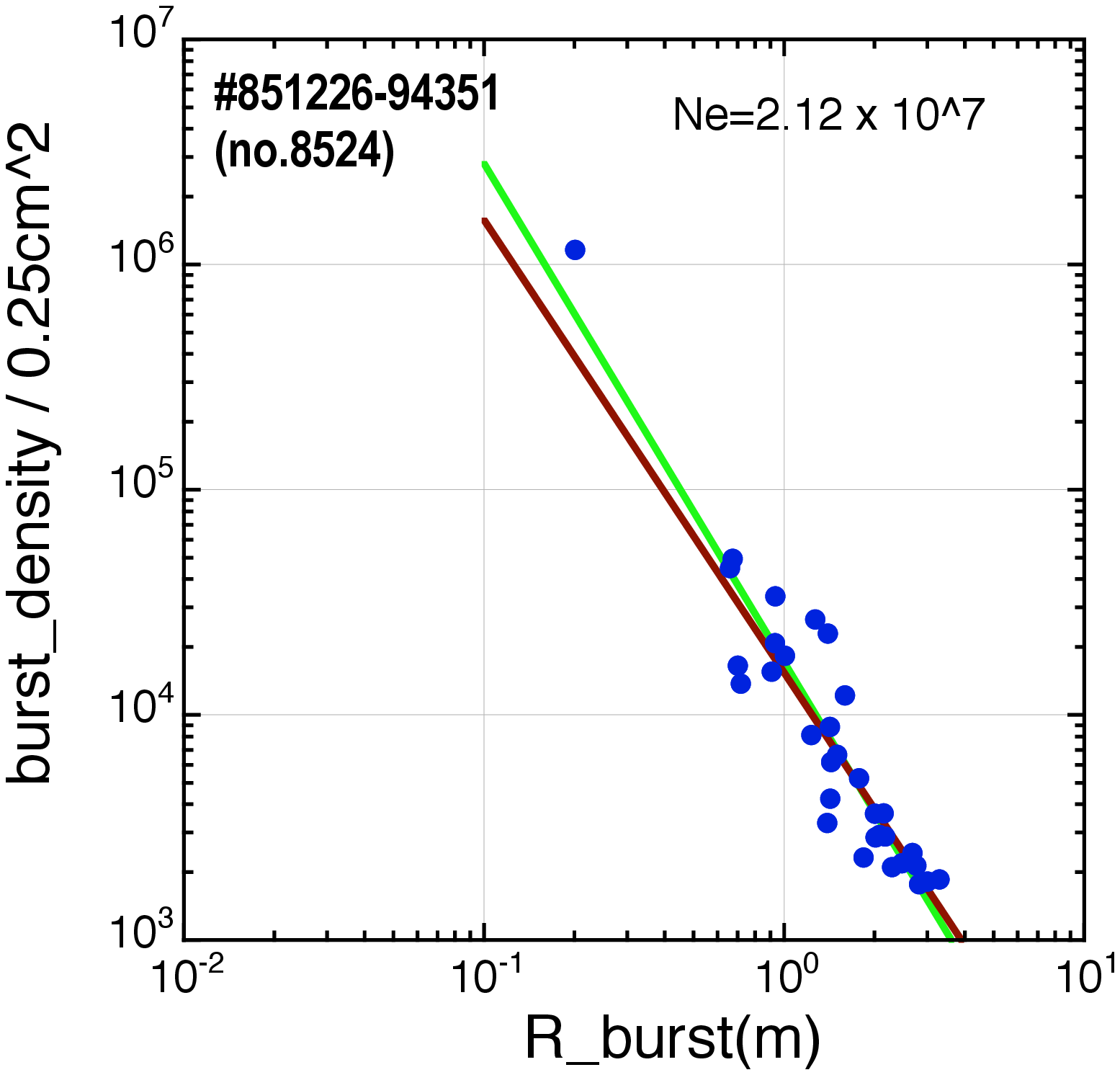}
\end{minipage} \hfill
\begin{minipage}{0.2\textwidth}
\hspace{-1cm}
\includegraphics[width=4cm]{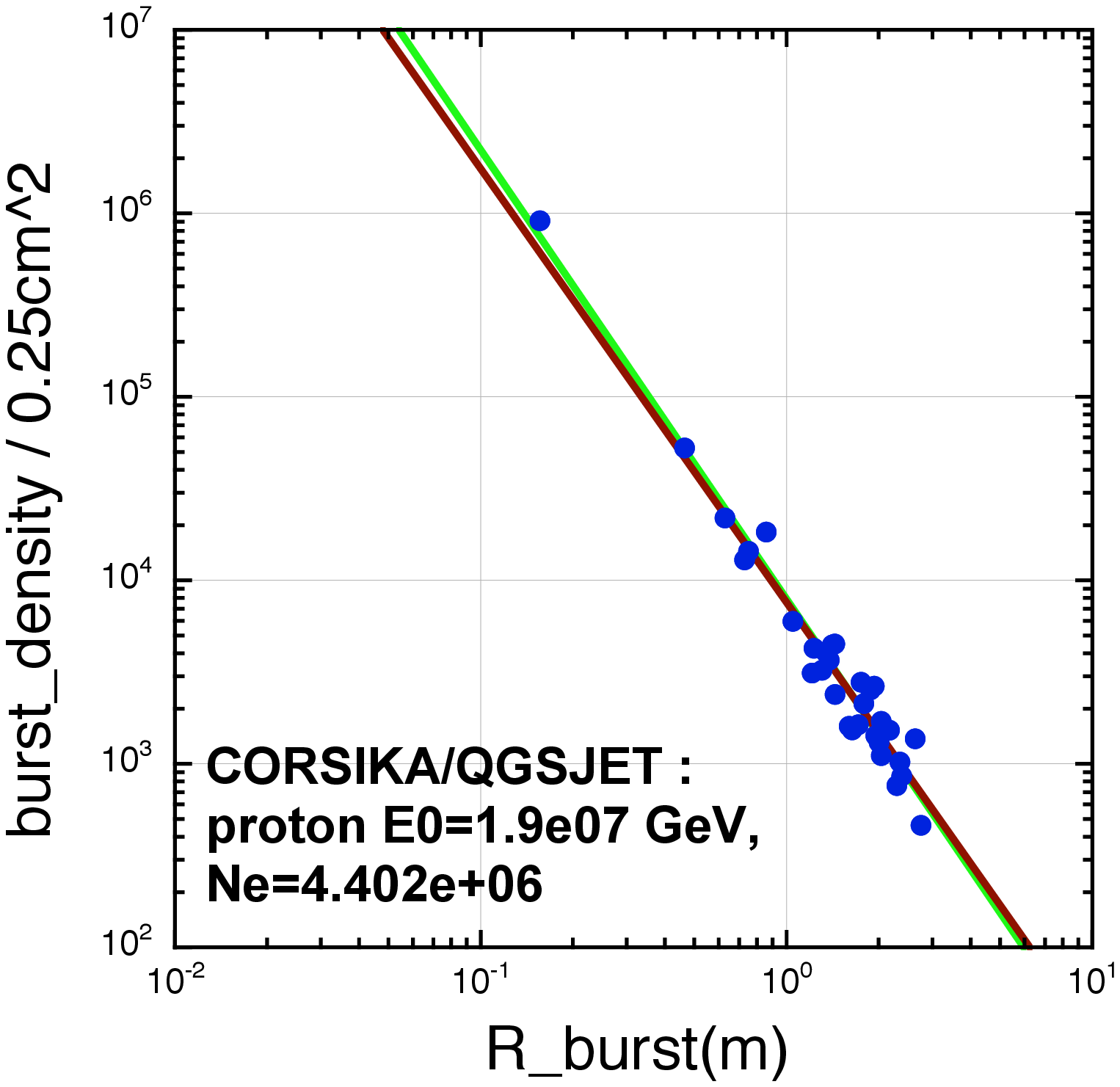}
\end{minipage}
\caption{\small  Examples of  burst data in the form of the lateral
distribution of burst-density for experimental data (left) and
for simulated data of proton-primary (right). }
\label{fig6}
\end{figure}

\subsection{High energy showers in emulsion chambers}
For high energy (e,$\gamma$)-particles and hadrons of 
$E \ge 1$ TeV in the atmospheric families 
arriving at  each emulsion chamber,  we calculate further the nuclear
and electromagnetic cascade development inside the chamber
taking into account  the exact structure of the emulsion chamber.
We use the QGSJET model for hadron-Pb interactions and a code
formulated by Okamoto and Shibata for electromagnetic
 cascades \cite{Shibata}.
 The electron number density under every 1cm Pb is transformed into
 spot darkness of the X-ray film.
Then the  energy of each shower is re-estimated from the shower transition
on spot darkness by applying the procedure
used in the experiments.
\subsection{Data of hadron calorimeters: Calculation of the burst-size}

Hadron calorimeters detect a bundle of charged particles, which are produced in the 
emulsion chamber material by the hadron component in the air-shower through  local
nuclear interactions. Output from each unit of the hadron calorimeter is
related to the energy deposited in the scintillator, and is converted to a
charged particle number using the average energy loss of a single muon in the
scintillator. The number of charged particles per 50 cm $\times$ 50 cm, $n_b$, is
called the ``burst density'' hereafter.
We use the GEANT4 code \cite{GEANT} for calculating the burst-density.
We calculate the average number of charged particles\footnote{
Here we take into account the scintillator response of charged particles.
Gamma-rays gives some energy deposit in the scintillator. Then the scintillator
response of gamma-rays are also taken into account\cite{Ohmori}.} 
  produced in the emulsion chamber of 15 cmPb 
  and arriving at the scintillator of the hadron calorimeter for  hadrons
 (pions, proton, kaons), muons and high energy $e,\gamma$ in the
air-shower,  with 4 different energies of 10 GeV, 100 GeV, 1 TeV and 10 TeV
and 5 different zenith angles of arrival direction.
Figure~\ref{fig5} shows an example of the energy dependence of average number of charged particles which responds
to scintillator for  six different incident particles.
The dependences are approximated by  numerical functions and extrapolate to higher
or lower energy range of the incident particles.
Applying these functions to every particle incident upon the emulsion chamber,
we get the burst-density in each block of 32 hadron calorimeters.
We define $n_b^{max}$ as the largest burst-density among the 32 blocks of the hadron calorimeters and 
$\Sigma n_b$ as the sum of burst-density of 32 blocks.
In the following we pick up events which satisfy the following criteria;
\begin{itemize}
\item $N_e \geq 10^6$,
\item $n_b^{max} \geq 10^4$,
\item $R_{AS-Bs} \le 1$m, \\ 
where $R_{AS-Bs}$ is the distance between the burst center and the air-shower center.
\end{itemize}
The burst center is determined by the algorithm described in Ref.\cite{SYS00}.
In the Chacaltaya data, 1,034 events satisfy the above criteria in $\sim 40$ m$^2$year
exposure of hadron calorimeters. 
Among them, 73 events are accompanied by high energy atmospheric families of
$\Sigma E_{\gamma} \geq$ 10 TeV ($E_{min} = 2$ TeV).
Figure~\ref{fig6} show examples of experimental and simulated burst data.
We can see the lateral distribution of the burst-density is well described by the power law
function \cite{SYS00}. 

\begin{figure*}
\begin{minipage}{0.3\textwidth}
\includegraphics[width=5.5cm]{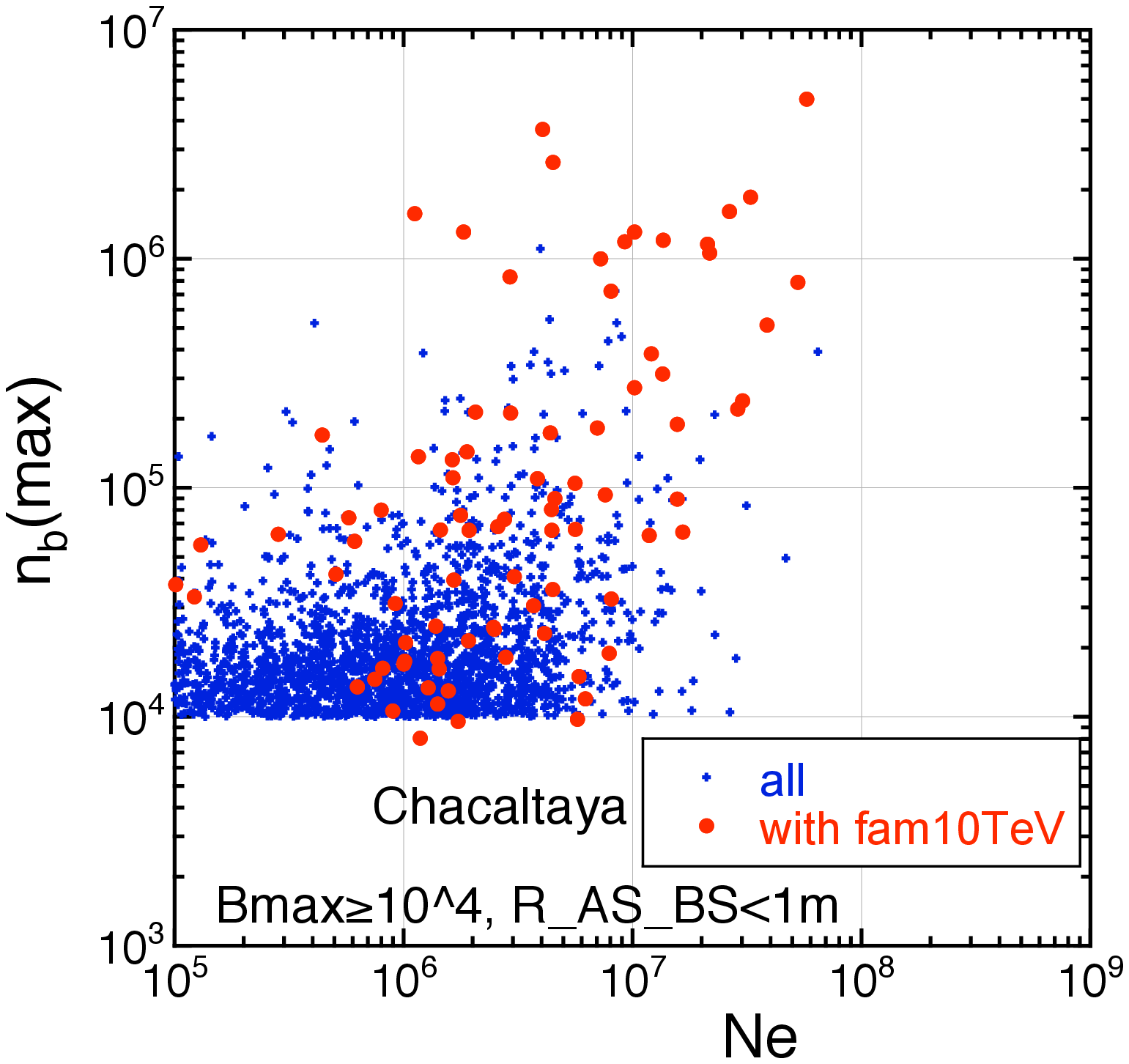}
\end{minipage} \hfill
\begin{minipage}{0.3\textwidth}
\includegraphics[width=5.5cm]{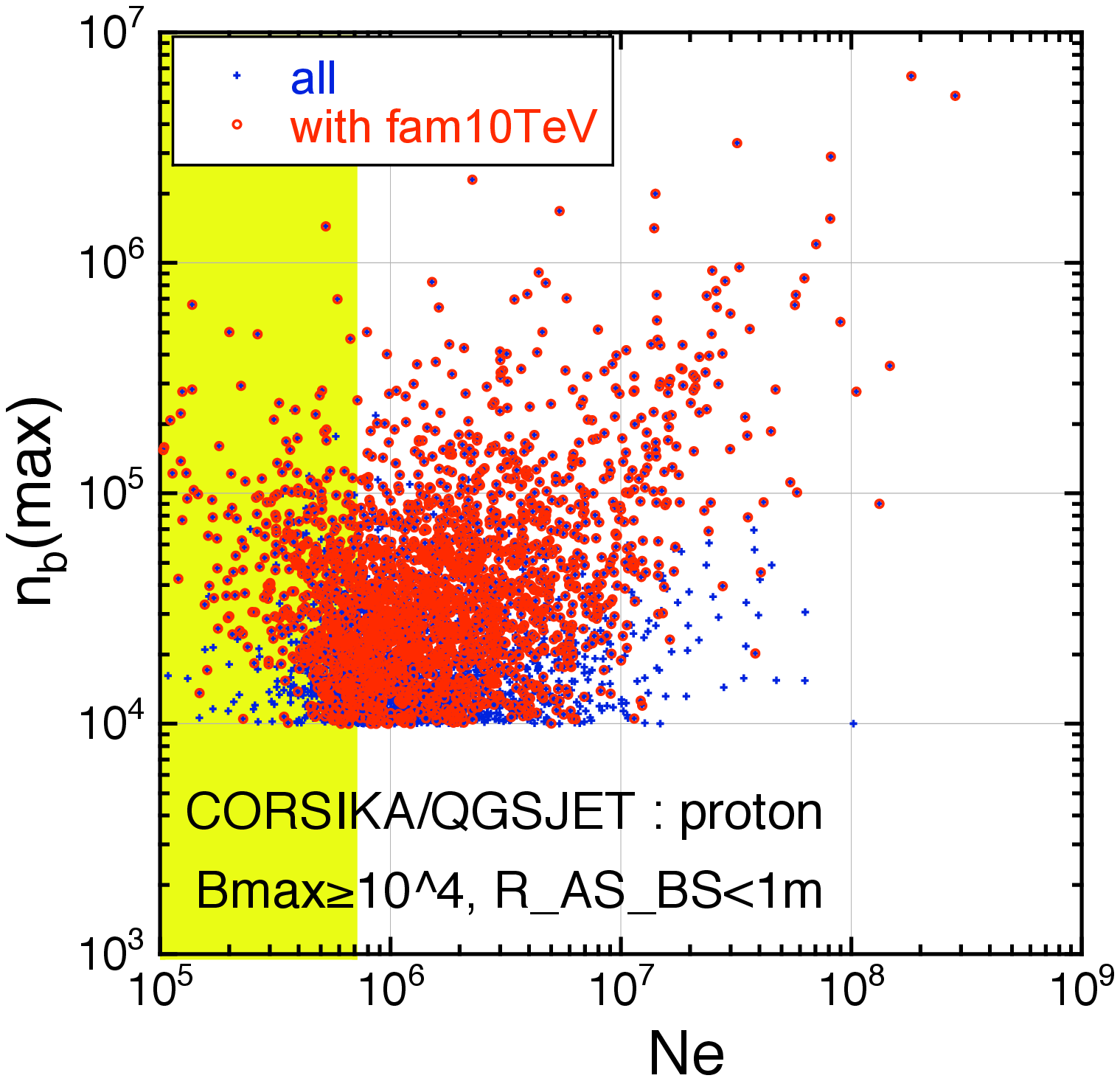}
\end{minipage} \hfill 
\begin{minipage}{0.3\textwidth}
\includegraphics[width=5.5cm]{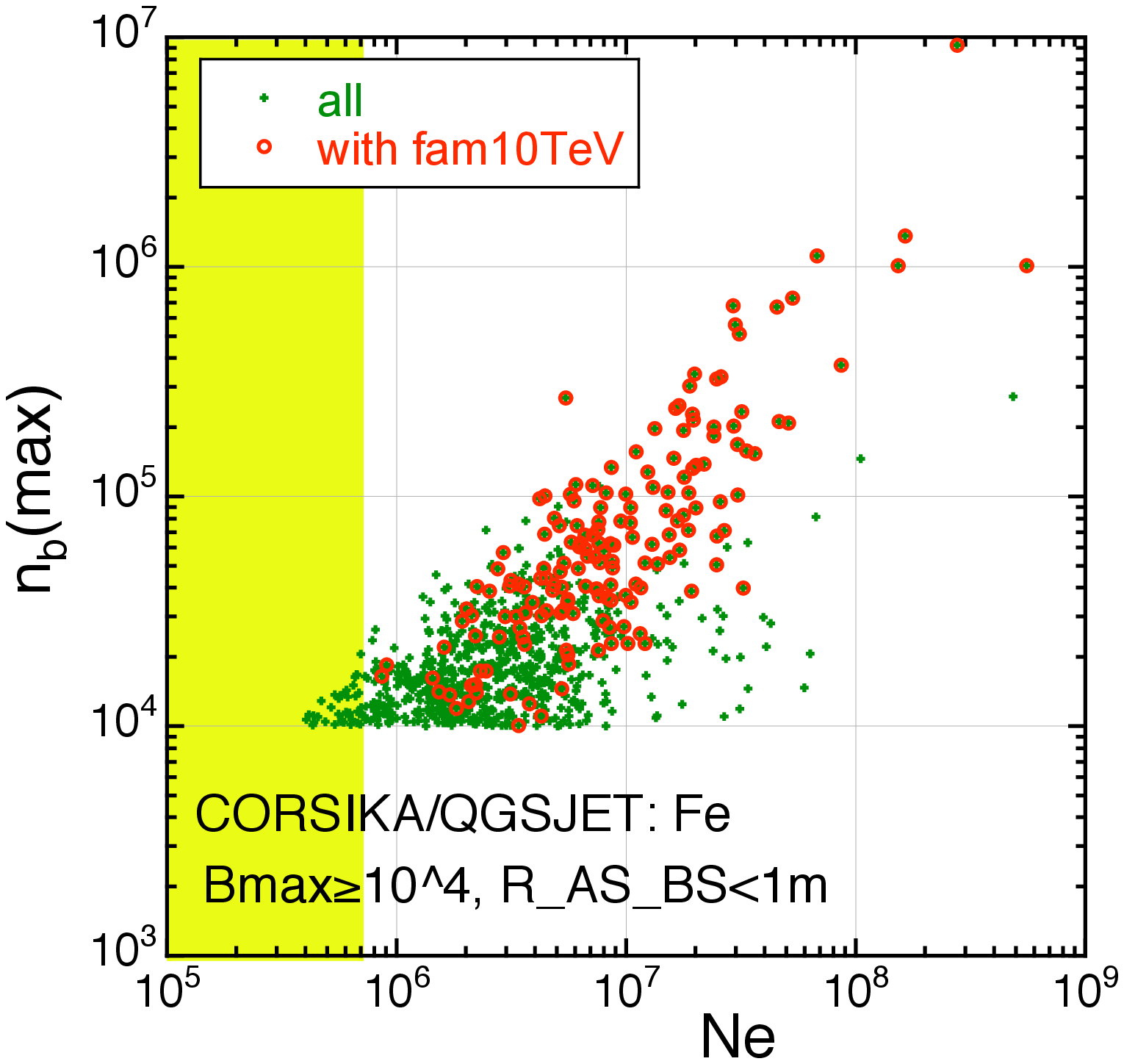}
\end{minipage}
\caption{\small  Scatter diagram between air-shower size, $N_e$, and maximum
burst density, $n_b^{max}$, of the event. Shaded area in the simulated data is biased
because the sampled primary energy is larger than $10^{15}$ eV. }
\label{fig7}
\end{figure*}

In Table 1, we show the fraction of proton, He, CNO, heavy and Fe components in the
air-showers accompanied by families and also those accompanied by burst.
In the shower-size region of $ N_e \ge 10^7$, corresponding to $E_0 \approx 10^{16}
- 10^{17}$ eV,  the fraction of each component is similar to those assumed in the primary
particles because almost all air-showers in this air-shower size region accompany families and bursts.

\section{Correlation between air-showers and bursts}

\begin{figure}
\begin{center}
\includegraphics[width=7cm]{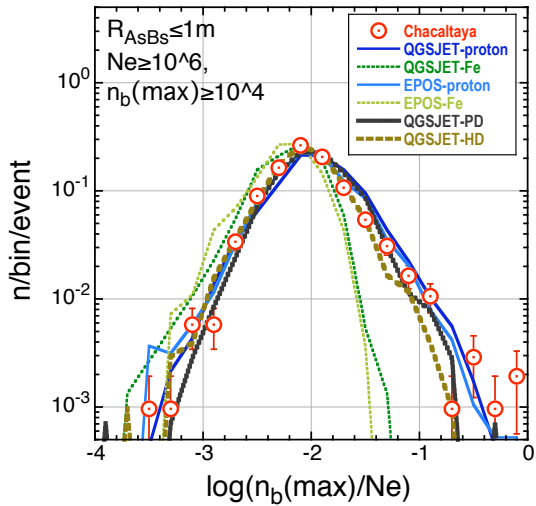}
\caption{\small 
Distribution of $n_b^{max}/Ne$. Circles are experimental data and
lines are simulated data, solid lines : proton-primaries,  dotted lines : Fe-primaries.
}
\label{fig8}
\end{center}
\end{figure}

\begin{figure}
\begin{center}
\includegraphics[width=6.8cm]{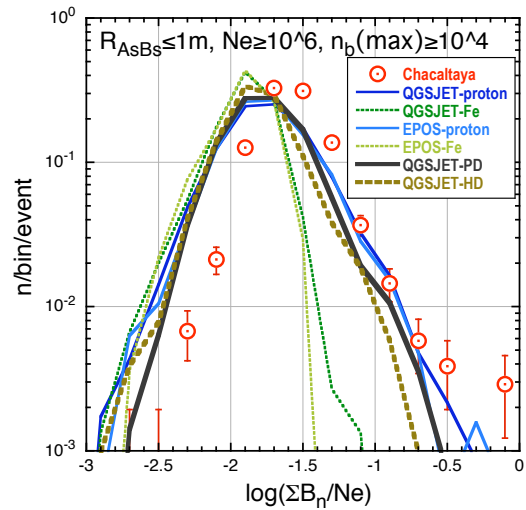}
\caption{\small 
Distribution of $\Sigma n_b/Ne$. Circles are experimental data and
lines are simulated data, solid lines : proton-primaries,  dotted lines : Fe-primaries.
}
\label{fig9}
\end{center}
\end{figure}

Figure~\ref{fig7} shows a scatter diagram between air-shower size, $N_e$, and maximum
burst density, $n_b^{max}$, of the event for the experimental data and for the
simulated data of proton- and Fe-primaries.
In the events of iron-primaries, $n_b^{max}$  is more or less proportional to $N_e$
though $n_b^{max}$ is weakly correlated to $N_e$ for the events of proton primaries.
It is very natural because Fe-air-nucleus interactions are assumed to be the superposition of a
number of low energy nucleon-air-nucleus collisions and so the fluctuation becomes small.
But for proton-air-nucleus interactions, the position of interactions and/or released energy at the
interaction fluctuate widely event by event.
The distribution in the experimental data looks close to that in proton-primaries.
Figure~\ref{fig8} shows distributions of $n_b^{max}/Ne$ for four different chemical composition
of primary particles, pure proton, pure iron, proton-dominant and heavy dominant.
The shape of the distribution for pure-iron primaries is very different from that for 
the others. 
There are almost no events with $n_b^{max}/Ne \ge \sim 0.02$  (log$(n_b^{max}/Ne) \ge \sim -1.6)$
in the iron-induced air-showers. On the contrary, a considerable number of events are found
in this region of the distribution for proton-induced air-showers.
There is no systematic difference in the shape for  the other three chemical compositions,
pure protons, proton-dominant and heavy dominant\footnote{
Nearly half of the air-showers accompanied by bursts are due to protons and He-nuclei,
even when heavy-dominant chemical composition is assumed in primary particles, as seen in Table 1. 
This is a reason why the difference in the shape of the distribution  
among these three chemical compostions is small.
}, and also for the two different interaction models,
QGSJET and EPOS.
The experimental data are well described by the model calculation for these three chemical
compositions of primary particles.
Figure~\ref{fig9} shows the distribution of $\Sigma n_b/Ne$ where $\Sigma n_b$ is a sum of $n_b$ over
32 blocks of hadron calorimeters.
Again we can see the experimental data are close to those expected for pure protons or
mixed chemical composition of proton-dominant and heavy-dominant,
though the number of events with smaller $\Sigma n_b$ are less than expectation.
An almost similar analysis was done by the Tibet group and they concluded that their data are well 
described by a heavy dominant composition (see Ref.\cite{Tibet00}).

\section{Correlation between bursts and families}

\begin{figure}
\begin{center}
\includegraphics[width=7cm]{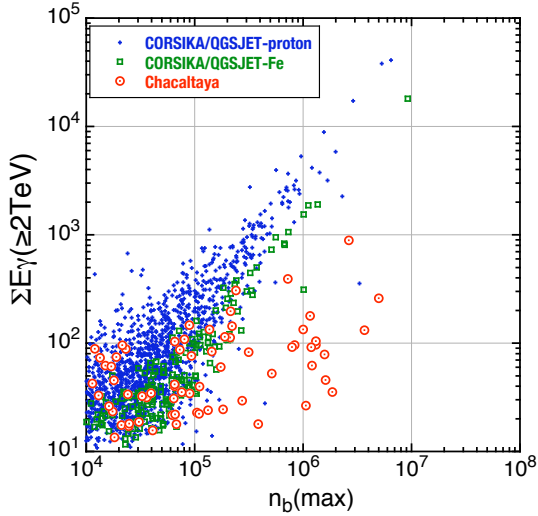}
\caption{\small 
Correlation diagram  between $n_b^{max}$ and family energy $\Sigma E_{\gamma}$
in the burst-triggered families in the air-showers of $N_e \ge 10^6$.
}
\label{fig10}
\end{center}
\end{figure}

\begin{figure}
\begin{center}
\includegraphics[width=7cm]{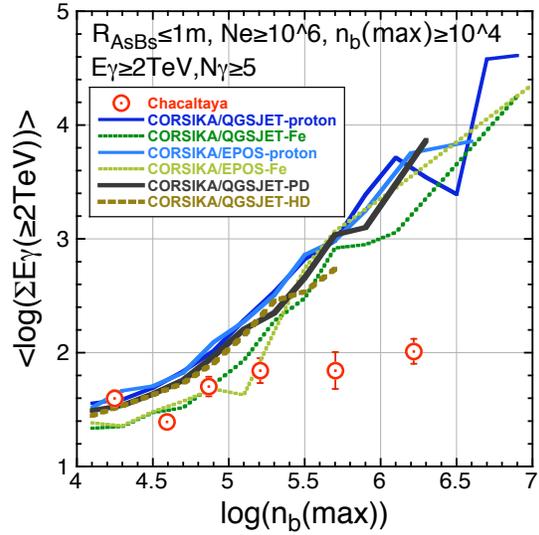}
\caption{\small 
Average family energy  versus $n_b^{max}$ in the burst-triggered families.
}
\label{fig11}
\end{center}
\end{figure}

Figure~\ref{fig10} shows a correlation diagram between $n_b^{max}$, maximum of burst size 
among 32 blocks of the event and accompanying family energy.
The experimental data are compared with those of simulated data of proton-primaries and 
iron-primaries. 
As seen in the figure, the family energy is almost proportional to $n_b^{max}$ in the simulated data
irrespective of the primary particles though
the family energy of the events coming from iron-primaries are smaller than that from proton-primaries.
In the figure we can see 
the family energy in the experimental data is systematically smaller than that
of simulated data in the events with larger burst-density, $n_b^{max} \ge 10^5$.
Figure~\ref{fig11} shows an average family energy versus $n_b^{max}$.
It is clear that the experimental data can not be explained simply by changing the chemical composition
of primary particles.

\begin{figure}
\begin{center}
\includegraphics[width=7cm]{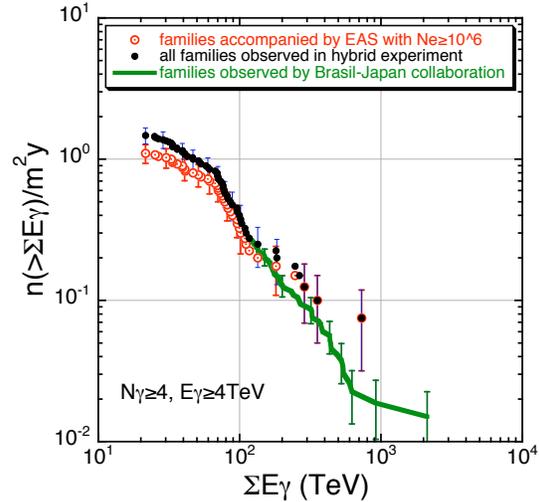}
\caption{\small 
Comparison of the integral family energy. The solid line is for the emulsion chamber
experiment of the Brazil-Japan Collaboration and solid circles are for air-shower
triggered families and open circles are for all families observed in the present
hybrid experiment.
}
\label{fig12}
\end{center}
\end{figure}

One may argue that the  smaller  family energy in the experimental data  is due to the systematic
underestimation of the shower energy in the emulsion chamber. 
Figure~\ref{fig12} shows a comparison of the integral spectra of family energy observed by the present 
hybrid experiment and  that by the  emulsion chamber
experiment of the Brazil-Japan collaboration\cite{Lattes80}.
As seen in the figure,  the two spectra smoothly connect with each other.
Thus we can conclude that the energy estimation of high energy particles in the emulsion chamber of the
present hybrid experiment is not much different from that of the emulsion chamber experiment of the 
Brazil-Japan collaboration. 

One may also argue about an overestimation of the burst-density, especially beyond the region of $n_b \ge 10^5$.
But this possibility is also ruled out, because  the distribution on $n_b^{max}/N_e$ is well described
 by the simulations as seen in Figure~\ref{fig8}.
\begin{figure}
\begin{center}
\includegraphics[width=7cm]{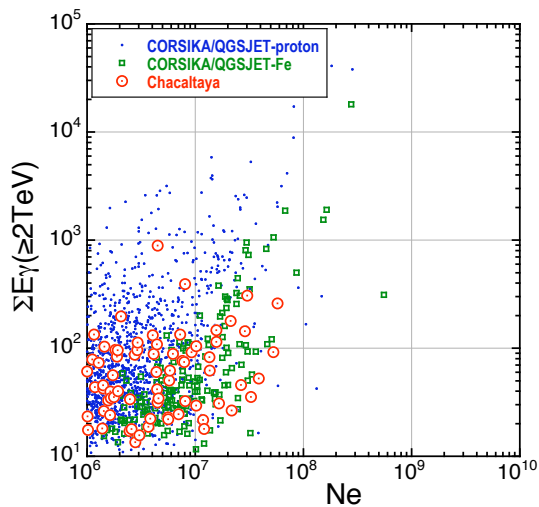}
\caption{\small 
Correlation diagram between air-shower size, $N_e$, and accompanied
family energy, $\Sigma E_{\gamma}$, of the events shown in Figure~\ref{fig10}.
}
\label{fig13}
\end{center}
\end{figure}

\section{Discussion}

The previous analysis on the air-shower-triggered families shows that the average family
energy of the experimental data is considerably smaller than that of simulations of proton-dominant primaries
 in the shower size region of $Ne \ge 10^7$, as described in section 2. 
 In Figure~\ref{fig13} we  show a correlation diagram between family energy, $\Sigma E_{\gamma}$
 and associated air-shower size, $N_e$, for the same events shown in Fig.10.
In the air-shower-size region of $N_e \ge 10^7$,  the family energy in the experimental data 
is systematically smaller than that expected for proton-primaries and those events look like coming
from iron-primaries.

The characteristics of the bursts accompanied by the proton-induced air-showers are very different from those
accompanied by iron-induced ones, as shown in Figure~\ref{fig7}.
The considerable number of air-showers induced by proton-primaries accompany large burst-density
which are not seen in the iron-induced air-showers, and the experimental data are close to those 
expected in case of proton-primaries.
The contradiction of the above two arguments is well seen in the correlation digram between bursts and
families, shown in Figure~\ref{fig10} and no model can describe the observed correlation.

The spectra of high energy particles detected by the emulsion chambers are more sensitive to the mechanism of
the particle  production of the most forward region of the rapidity. 
The hadron calorimeter supplies the data of hadron component in the air-shower.
The hadron component in the air-shower bears more direct information on the
nuclear interaction than any other component such as electron  and muon component
in the air-shower, and the burst-size of hadron calorimeter gives a measure of interaction energy.

Suppose the $x$-distribution of the produced particles becomes steeper, the number of 
high energy particles detected in the emulsion chamber becomes small
 (i.e., detected family energy becomes smaller),
because of the high threshold energy of the emulsion chambers.
The hadron component detected by hadron calorimeters, however, does not 
change much because of the lower detection threshold energy.
Then the ration of $\Sigma E_{\gamma}/n_b^{max}$ becomes smaller.
The observed discrepancy between experimental data and simulated data can be explain in this way,
that is, the experimental data indicate the $x$-distribution of produced particles in the
cosmic-ray nuclear interactions of $E_0 \ge \sim 10^{16}$ eV is much steeper than that 
assumed in the models.

\vspace{1cm}

\begin{acknowledgments}

The author wishes to thank all the member of Chacaltaya hybrid experimemt,  
especially to Drs. N.Kawasumi, K.Honda,  N.Ohmori, N.Inoue
and A.Ohsawa  for their valuable comments and discussions on the details of
experimental procedures.

\end{acknowledgments}



\begin{thebibliography}{99}

\bibitem{SYS96} 
  N.Kawasumi et al.,
  Phys. Rev. D {\bf 53} (1996) 3534
\bibitem{SYS00} 
  C.Aguirre et al.,
  Phys. Rev. D {\bf 62} (2000) 032003
\bibitem{Tibet00} 
  Tibet AS$\gamma$ Collaboration (M.Amenomori et al.),
  Phys. Rev. D {\bf 62} (2000) 112002-1, 072007-3
\bibitem{TienShan} 
  S.B.Shaulov,
  AIP Conf. Proc. {\bf 276} (1992) 94
\bibitem{BASJE}
  Y.Shirasaki et al.,
  Astropart. Phys. {\bf 15} (2001) 357
\bibitem{Kascade}
  T.Antoni et al. KASCADE Collboration,
  Astropart. Phys. {\bf 14} (2001) 245, {\bf 19} (2003) 703,715 
\bibitem{AP18_Swordy}
  S.P.Swordy et al.,
  Astrop. Phys. {\bf 18} (2002) 129
\bibitem{Tibet06} 
  Tibet AS$\gamma$ Collaboration (M.Amenomori et al.),
  Phys. Lett. B {\bf 632} (2006) 58
\bibitem{EPOS06}
  K.Werner, F.M.Liu and T.Pierog,
  Phys. Rev. C {\bf 74} (2006) 044902
\bibitem{EPOS07}
  T.Pierog and K.Werner,
  Proceedings of 30th ICRC, Merida (2007), Vol.4, p.629
\bibitem{H.Ulrich}
  H.Urlich et al.,
  Proceedings of 30th ICRC, Merida (2007), Vol.4, p.87
\bibitem{SYS_Merida}
  H.Aoki et al.,
  Proceedings of 30th ICRC, Merida (2007), Vol.4, p.23
\bibitem{icrc09_tama}
S.P.Besshapov et al.,
Nucl. Phys. B (Proc. Suppl.) 196(2009) 118;~
  Proceedings of 31th ICRC, Lodz (2009)  \#0214
\bibitem{Horandel03}
  J.R.Horandel,
  Astropart. Phys. {\bf 19} (2003) 193
\bibitem{CORSIKA} 
  D.Heck, J.Knapp, J.N.Capdevielle, G.Schatz and T.Thouw,
  Fortshungzentrum Karlsruhe, FZKA 6019 (1998)
\bibitem{QGSJET} 
  N.N.Kalmykov and S.S.Ostapchenko,
  Yad. Fiz. {\bf 56} (1993) 105
\bibitem{Sibyll94} 
  R.S.Fletcher et al.,
  Phys. Rev. {\bf D50} (1994) 5710
\bibitem{Sibyll92}
  J.Engel et al.,
  Phys. Rev. {\bf D46} (1992) 5013,
\bibitem{Shibata} 
  M.Okamoto and T.Shibata,
  Nucl. Instr. and Meth. A {\bf 257} (1987) 155
\bibitem{GEANT}
S.Agostinlli et al. Geant4 collaboration, Nucl. Inst. and Meth. A506(2003) 250
\bibitem{Ohmori}
N.Ohmori, private communication.
\bibitem{Lattes80}
C.M.G.Lattes, Y.Fujimoto and S.Hasegawa, Phys. Rep. Vol.65 (1980) 151
\end{thebibliography}
\end{document}